# The response of pre-osteoblasts and osteoclasts to gallium containing mesoporous bioactive glasses

N. Gómez-Cerezo[1,2], E. Verron[3], V. Montouillout[4], F. Fayon[4], P. Lagadec[5], J.M. Bouler[3], B. Bujoli[3], D. Arcos[1,2,*], M. Vallet-Regí[1,2,*].

1. Departamento de Química en Ciencias Farmacéuticas, Facultad de Farmacia, Universidad Complutense de Madrid, Instituto de Investigación Sanitaria Hospital 12 de Octubre i+12, Plaza Ramón y Cajal s/n, 28040 Madrid, Spain

2. CIBER de Bioingeniería Biomateriales y Nanomedicina (CIBER-BBN), Spain

3. Université de Nantes, CNRS, UMR 6230, CEISAM, UFR Sciences et Techniques, 2 Rue de la Houssinière, 44322 NANTES Cedex 3, France.

4. CNRS, UPR 3079, CEMHTI, 1D Avenue de la Recherche Scientifique, 45071 Orléans Cedex 02, France.

5. Université Côte d'Azur, CNRS, Inserm, institut de Biologie Valrose (iBV), 28 Av. de Valombrose, 06107 Nice cedex 2, France

*Corresponding authors

E-mail address : arcosd@ucm.es (Daniel Arcos), vallet@ucm.es (María Vallet-Regí)

**Abstract**

Mesoporous bioactive glasses (MBGs) in the system $SiO_2$-$CaO$-$P_2O_5$-$Ga_2O_3$ have been synthesized by the evaporation induced self-assembly method and subsequent impregnation with Ga cations. Two different compositions have been prepared and the local environment of Ga(III) has been characterized using $^{29}Si$, $^{71}Ga$ and $^{31}P$ NMR analysis, demonstrating that Ga(III) is efficiently incorporated as both, network former ($GaO_4$ units) and network modifier ($GaO_6$ units). *In vitro* bioactivity tests evidenced that Ga-containing MBGs retain their capability for nucleation and growth of an apatite-like layer in contact with a simulated body fluid with ion concentrations nearly equal to those of human blood plasma. Finally, *in vitro* cell culture tests evidenced that Ga incorporation results in a selective effect on osteoblasts and osteoclasts. Indeed, the presence of this element enhances the early differentiation towards osteoblast phenotype while disturbing osteoclastogenesis. Considering these results, Ga-doped MBGs might be proposed as bone substitutes, especially in osteoporosis scenarios.

**Keywords:** gallium, mesoporous bioactive glasses, osteoblast, osteoclast, osteoporosis

# 1. Introduction

Mesoporous bioactive glasses (MBGs) are a kind of bioceramics with potential application for bone tissue regeneration purposes. Since Yan et al prepared MBGs for the first time in 2004[1], these materials have arisen great interest in the field of regenerative medicine of bone due to their excellent bioactive properties, which have been demonstrated both in vitro [2,3] and in vivo[4,5]. MBGs are commonly synthesized in the ternary $SiO_2$-$CaO$-$P_2O_5$ system, with similar chemical compositions to those used for conventional bioactive sol-gel glasses [2,6,7]. However, the incorporation of a structure directing agent during the hydrolysis and condensation processes leads to the formation of a mesophase, which eventually results into ordered mesoporous structures with very high surface areas and porosities[8]. The bone regeneration properties of bioglasses come up from their capability to exchange ions with the surrounding fluids, mainly $Ca^{2+}$ by $H^+$, with the subsequent growth of an apatite-like phase on the surface, similar to the mineral component of bones and teeth[9]. Besides, the dissolution products of bioactive glasses ($Ca^{2+}$ and silica species in solution) elicit the upregulation of pro-osteoblastic genes [10–12] thus enhancing their osteogenic potential.

The high surface area and porosity of MBGs accelerate the chemical reactions at the implants surface and, consequently, MBGs exhibit the fastest bioactive behaviour observed so far in the field of bioceramics [13]. The ordered mesoporous structure of MBGs brings about their potential applications as drug delivery systems [14] and are excellent matrixes to host and release active agents such as antibiotics [15], antiosteoporotics [16] or antitumoral drugs [17], which can even be released in a stimuli-responsive manner [18,19].

The efficiency of MBGs to trigger the ionic exchange with the surrounding fluids has recently fostered their application as carriers of therapeutic ions. In this way species, such as $Li^+$, $Sr^{2+}$, $Cu^{2+}$ $Co^{2+}$ or $B^{3+}$ have been incorporated into $SiO_2$-$CaO$-$P_2O_5$ MBGs, with the aim of providing osteogenic, cementogenic, angiogenic and antibacterial properties [20] [21–24]. The release of these chemical species can yield very interesting synergies with the bioactive behaviour of MBGs and the pharmacological treatment loaded in the mesopores. Therapeutic ions are commonly incorporated to MBGs as inorganic salts dissolved together with the MBGs precursors. At the end of the synthesis these ions can stay in the MBGs associated to phosphates or as network modifiers of the silica matrix, similarly to the role played by calcium cations [25–29] . In most cases, the large volume and/or the polarization capacity of these ions are so intense that even low concentrations lead to the disorganization of the mesoporous structure [30].

One of the most interesting therapeutic ions in the field of bone pathologies is Ga (III). The inhibition of bone resorption driven by Ga (III) has been widely reported [31–35]. Ga reduces the resorption activity, differentiation and formation of osteoclasts without cytotoxic effects in a dose-dependent manner, pointing out an excellent potential for the treatment of osteoporotic bone fractures associated to bone implants [36,37]. In this sense, Ga ions have been successfully combined with apatitic calcium phosphate cements and tested both in vitro and *in vivo* [38].

Although some studies have been carried out describing the effects of $Ga^{3+}$ on the structure of MBGs [39–42], as well as haemostatic and antibacterial effects [43] , from the best of our knowledge no study has considered the effects of Ga containing MBGs on bone cells, such as osteoblasts or osteoclasts. This is quite surprising considering the very interesting synergy between $Ga^{3+}$ and MBGs for the treatment of osteoporotic bone fractures. Moreover, contrarily to monovalent or divalent Cu(II), Co (II), Sr (II), Li (I), etc. cations, which act as modifiers of the silica network when incorporated into MBGs, Ga (III) cations can act as both network formers (NF) and network modifier (NM). The distribution and local environment of Ga species into MBGs are still poorly understood and the influence of Ga on bone remodelling cannot be explained without understanding the relationship between the local structure of this inhibitor of bone resorption with the resulting biological response.

In this paper we evaluate for the first time the biological response of Ga doped MBGs with respect to bone forming cells (osteoblasts) and bone resorptive cells (osteoclasts), together with a deep study of the local environment of this cation into the mesoporous matrix of these bioceramics.

## 2. Materials and methods

### 2.1 Synthesis of materials

Two different MBGs in the system $SiO_2$-$CaO$-$P_2O_5$ were synthetized by evaporation induced self-assembly (EISA) method [44] . For this aim, triblock copolymer F127 $(EO)_{100}$-$(PO)_{65}$-$(EO)$ was used as structure directing agent. Tetraethyl orthosilicate (TEOS) (98%), triethyl phosphate (TEP) (99%) and calcium nitrate tetrahydrate $Ca(NO_3)_2 \cdot 4H_2O$ (99%) were used as $SiO_2$, $P_2O_5$ and CaO sources, respectively. In a typical synthesis 2 g of surfactant were dissolved in ethanol with HCl 0,5 M solution at room temperature. Afterward TEOS, TEP and $Ca(NO_3)_2 \cdot 4H_2O$ were added under stirring in 3 hours intervals to obtain two different MBG denoted MBG-58S and MBG-85S. The different reagents and amounts are shown in Table 1. The resulting solution was kept under stirring for 12 hours and poured into Petri dishes (9 cm in diameter). The colourless solution was evaporated at 30ºC for 7 days. Eventually, the dried gels were removed as homogeneous and transparent membranes and calcined at 700ºC for 3 hours under air atmosphere.

Solids MBG-58SGa and MBG-85SGa were synthetized from MBG-58S and MBG-85S respectively. In a typical procedure, a certain amount of $Ga(NO_3)_3 \cdot 4H_2O$ was dissolved in a vessel containing 100 mL of ultrapure water. Then 500 mg of the corresponding MBG (MBG-58S or MBG-85S) were added to the solution in such a way that the Ga/Ca molar ratio was 0.38 (see Table 1). The pH of the solution was adjusted in the 9.2-9.8 range by means of a concentrated ammonia solution. The mixture was stirred for 24 hours at room temperature and the solids obtained were filtered under vacuum. Finally, solids were dried at 100ºC for 24 hours. All reactants were purchased from Aldrich and used without further purification.

## 2.2 Physicochemical characterization of MBGs

Low angle powder X-ray diffraction experiments were performed with a Philips X'Pert diffractometer equipped with Cu Kα radiation (wavelength 1.5406 Å). XRD patterns were collected in the 2θ° range between 0,5 and 6,5 with a step size of 0.02 2θ° and counting time of 4s per step.

The textural properties of the calcined materials were determined by nitrogen adsorption with a Micromeritics ASAP 2010 equipment. To perform the $N_2$ adsorption measurements, the samples were previously degassed under vacuum for 20 h, at 100 ºC. The surface area was determined using the Brunauer-Emmett-Teller (BET) method. The pore size distribution between 0.5 and 40 nm was determined from the adsorption branch of the isotherm by means of the Barret-Joyner-Halenda (BJH) method. This experiment was made in triplicate using different batches of each sample.

Chemical compositions of MBGs were determined by X-ray fluorescence (XRF) spectroscopy, using a Philips PANalytical AXIOS spectrometer (Philips Electronics NV), with X-rays generated by the RhKα line at λ = 0.614 Å.

$^{29}$Si MAS and {$^{1}$H}-$^{29}$Si CP-MAS NMR spectra were recorded on a Bruker Avance 400WB spectrometer operating at a magnetic field of 9.4 T ($^{29}$Si Larmor frequency of 79.5MHz) at a spinning frequency of 12 kHz. The 29Si quantitative MAS spectra were acquired using a 90° flip angle with a recycling delay of 5s and around 10000 scans were co-added. The {$^{1}$H}-$^{29}$Si CP-MAS spectra were recorded using a contact time of 3.5 ms. 135000 transients were co-added with a recycle delay of 5s. The $^{71}$Ga NMR experiments were performed on a Bruker Avance III spectrometer operating at 20.0 T ($^{71}$Ga Larmor frequency of 259.3MHz) equipped with a 1.3mm double-resonance MAS probe head. The $^{71}$Ga central transition (CT)-selective MAS spectra were recorded using a Hahn echo sequence with whole echo acquisition. The echo delay was synchronized with the rotor period and the rf-field strength was 60 kHz corresponding to a CT-selective π/2 pulse of 2.1 μs. Approximately, 28 000 transients were accumulated for MBG-58SGa and 122 500 for MBG-85SGa with 0.5 s recycling delay. The $^{31}$P quantitative MAS spectra were recorded using single pulse (π/10)

acquisition; ~500 scans were co-added with a recycle delay of 120 s to ensure complete recovering of the longitudinal magnetization. $^1$H SPINAL-64 decoupling with a nutation frequency of 70 kHz was applied during signal acquisition[45]. Chemical shift values were referenced to tetramethylsilane (TMS), $H_3PO_4$ and $Ga(NO_3)_3 \cdot 4H_2O$ for $^{29}$Si, $^{31}$P and $^{71}$Ga, respectively

Transmission electron microscopy (TEM) was carried out using a JEOL-2100 microscope, operating at 300 kV (Cs 0.6 mm, resolution 1.7 Å). Images were recorded using a CCD camera (model Keen view, SIS analyses size 1024 x 1024, pixel size 23.5 mm x 23.5mm) at 30000x and 60000x magnification using a low-dose condition.

### 2.3 *In vitro* bioactivity tests

Assessments of *in vitro* bioactivity were carried out on powder samples. For this purpose, 50 mg of powder were soaked into 7 mL of filtered simulated body fluid (SBF), in polyethylene containers at 37 ºC under sterile conditions. SBF solution was prepared according to Kokubo et al. [46] by dissolving NaCl, KCl, $NaHCO_3$, $K_2HPO_4 \cdot 3H_2O$, $MgCl_2 \cdot 6H_2O$, $CaCl_2$, and $Na_2SO_4$ into distilled water. It was buffered at pH = 7.38 by using tris(hydroxymethyl)-aminomethane/HCl (TRIS/HCl) and then passed through 0.22 μm Millipore filters to avoid bacterial contamination. To avoid false positive responses, we included α-$Al_2O_3$ as inert control (Fig S5 in supporting information). The evolution of the glass surface was analysed by Fourier transform infrared (FTIR) spectroscopy with a Nicolet Magma IR 550 spectrometer and by scanning electron microscopy (SEM) using a JEOL F-6335 microscope. To assess the $Ca^{2+}$-$H^+$ ionic exchange between MBGs and SFB, calcium concentration and pH changes were analysed as a function of soaking time by using an Ilyte $Na^+$ $K^+$ $Ca^{2+}$ pH system. All the experiments were carried out in triplicate

### 2.4 Ga *in vitro* release

Ga release study was performed suspending 2 mg of solids in 2 mL of α-MEM (Sigma Chemical Company, St. Louis, MO, USA). 0.5 mL of this suspension was placed on a transwell permeable support with polycarbonate membrane (0.4 lm). The well was filled with 1.5 mL of αMEM and the suspension was stirred at 37 C at 100 rpm for 10 days. The amount of Ga released was determined by inductively coupled plasma spectroscopy (ICP), and the solution outside the transwell insert was replaced with fresh medium after each measurement. All the experiments were carried out in triplicate.

**2.5 Pre-osteoblast cell culture tests**

Murine MC3T3-E1 pre-osteoblasts ($2·10^4$ cell/mL) were cultured in α-MEM with 10% fetal bovine serum (FBS, Gibco, BRL), 1 mM L-glutamine (BioWhittaker Europe, Belgium), penicillin (200 mg·mL$^{-1}$, BioWhittaker Europe, Belgium), and streptomycin (200 mg·mL$^{-1}$), BioWhittaker Europe, Belgium), under a $CO_2$ (5%) atmosphere at 37 ºC for 12 hours to reach the confluence in each cell plate. After that the MBGs powders were added on the cells at 1mg/ml of concentration.

*2.5.1 Fluorescence microscopy assays.*

To evaluate cells morphology after 4 days in contact with the materials, MC3T3-E1 cells were cultured in 24 well plates at density of $10^4$ cells/mL in the presence of 1mg/mL of the different MBGs. Thereafter, cells were rinsed with PBS twice and fixed in 4% (w/v) paraformaldehyde in PBS for 2 hours. The samples were incubated 15 min with Atto 565-conjugated phalloidin (dilution 1:40, Molecular Probes) which stains actin filaments. Samples were then washed with PBS and the cell nuclei were stained with lM40-6 diamino-20-phenylindole in PBS (DAPI) (Molecular Probes). Fluorescence microscopy was performed with an Evos FL Cell Imaging System equipped with tree Led Lights Cubes (lEX (nm); lEM (nm)): DAPY (357/44; 447/60), GFP (470/22; 525/50), RFP (531/40; 593/40) from AMG (Advance Microscopy Group). Red channel was used to evaluate the cytoskeleton and blue for cell nucleus. All the experiments were carried out in triplicate.

*2.5.2 Pre-osteoblast proliferation studies*

The proliferation was determined by means of the Alamarblue® method at 1, 4, and 7 days of culture in contact with the different samples. The Alamarblue® method is based on the reduction of blue fluorogen (resazurin) to a red fluorescent compound (resofurin) by intracellular redox enzymes. A solution of resazurin sodium salt (Sigma Aldrich) at 0.01 mg/mL was prepared in phosphate-buffered saline (PBS) and then diluted 1:10 with culture medium supplemented with 10 % FBS. The cells were exposed to resazurin solution for 3 h at 37° C under $CO_2$ (5 %) atmosphere. Then fluorescence signal was read at $\lambda_{em}$=590 nm using a $\lambda_{exc}$= 560nm with a fluorescence spectrometer Biotek Synergy 4 device. Thereafter, the medium was renewed to continue the cell proliferation measurements at different times. Polystyrene culture plate was used as control and all the experiments were carried out in n = 4.

*2.5.3 Pre-Osteoblasts early differentiation studies*

The alkaline phosphatase (ALP) activity after 7 days in the presence of 1mg/mL of the MBGs, was measured to evaluate the influence of the Ga as a differentiation marker in assessing the expression of the osteoblast phenotype. For this purpose, $3 \cdot 10^4$ MC3T3-E1 cells were seeded with a 1 mg/mL using supplemented medium with β-glycerolphosphate (50 mg mL$^{-1}$, Sigma Chemical Company, St. Louis, MO, USA) and L-ascorbic acid (10 mM, Sigma Chemical Company, St. Louis, MO, USA). Solids were irrigated with PBS two times to remove as much residual serum possible. The cells were detached by subjecting the plates to cycles of freezing and thawing before measuring the ALP activity and total protein content. ALP activity was measured based on the hydrolysis of p-nitrophenylphosphate to p-nitrophenol. After 30 min incubation at 37ºC the reaction was stopped by the addition of 125 μL of 2M NaOH. The solution obtained was measured using a Helios Zeta UV-VIS spectrophotometer at 410 nm. Total protein content was determined using a colorimetric method at 540 nm (Bio-Analítica,S.L.) with a Helios Zeta UV-VIS spectrophotometer. Polystyrene culture plate was used as control and all the experiments were carried out in triplicate.

**2.6 Osteoclasts culture tests**

The mouse monocyte cell line RAW 264.7 was obtained from ATCC (Ref. # TIB-71; LGC Standards, Molsheim, France). Cells were cultured in Dulbecco's Modified Eagle's Medium (DMEM, Gibco, Paisley, UK) containing 5% Hyclone serum (Thermo Scientific Braunschweig, Germany) and 1% penicillin/streptomycin (Gibco, Paisley, UK). For osteoclast differentiation experiments, RAW 264.7 cells were seeded at 5,000 cells/cm$^2$ in alpha Minimum Essential Medium (α-MEM, Gibco, Paisley, UK) containing 5% Hyclone serum and effectors were added immediately. RANKL (Receptor Activator of Nuclear Factor- B Ligand) was used at 20 nM. Mouse effector GST-RANKL was produced as described by Strazic et al [47], and a GST protein was produced and purified using the same protocol and was used as a control. Cells were cultured for four days with a renewal of the medium at day 2.

*2.6.1.Viability assay*

Raw 264.7 cells were seeded at a density of 5000 cells/cm$^2$ and treated with the different MBGs at 10mg/mL or culture medium as control. After 4 days of incubation, the cells were incubated in a solution of 3-(4.5-dimethylthiazol-2-yl)-2.5-diphenyltetrazoliumbromide (MTT) at 0.5mg/mL for 4h. Then, the MTT solution was removed and replace by the same amount of the eluting buffer (DMSO/Ethanol 50:50). After 2h agitation on an orbital shaker, 80µl of the colourful supernatant were transferred into a new 96-well plate which was read at 562 nm.

*2.6.2.Tartrate-resistant acid phosphatase (ACP5) staining*
Raw 264.7 cells were seeded as previously described and treated with the different types of MBGs (10mg/mL) for 4 days. For control, cells were cultured without MBG. After 4 days, cells were rinsed two times in PBS and fixed in citrate/acetone solution for 30 s. ACP5$^+$ cells were revealed using the Acid Phosphatase, Leukocyte kit following manufacturer's instructions (# 387A; Sigma-Aldrich, USA). Stained cells were observed using a light microscope (Axioplan 2, Zeiss, Germany) and the number of TRAP-positive multinucleated cells was evaluated by comparing 8 fields between the different treatments. To further quantify the TRAP staining, DMSO was added, and the plate was centrifuged for 15 min at 150 rpm at room temperature in dark. Finally, optical density at 562 nm was read [47].

**Statistical analysis**

The statistics data are expressed as means-standard deviations of experiments. Statistical analysis was performed using the Statistical Package for the Social Sciences (SPSS) version 11.5 software. Statistical comparisons were made by analysis of variance (ANOVA). Subsequently, post hoc analyses were carried out to correct for multiple comparisons. In all the statistical evaluations, $P < 0.05$ was considered as statistically significant.

## 3. Results

### 3.1 Physicochemical characterization of MBGs

The mesoporous structure of the MBGs was studied by XRD and TEM (Figure 1). Both studies showed a disordered mesoporous structure for MBG-58S and MBG-58SGa materials, as usually found for MBGs with high calcium content. On the contrary, MBG-85S and MBG-85SGa exhibit a highly ordered hexagonal structure with planar group *p6mm* and parallel channels with a d-spacing of 5-6 nm, similarly to the silica mesoporous material SBA-15 [48]. To obtain a deeper knowledge of the mesoporous structure, textural properties were analyzed by $N_2$ adsorption/desorption method. All the curves could be identified as type IV isotherms characteristic of mesoporous materials (Figure S1 in supporting information). Surface areas, pore volumes and pore sizes of the four MBGs are collected in Table 2. Gallium-free MBGs show surface areas and pore volumes of in the range commonly observed for MBGs when prepared with F127 as structure directing agent [8]. After Ga incorporation, a decrease of surface area and pore volume is observed for both compositions (MBG-58SGa and MBG-85SGa). Regarding pore size, Ga incorporation comprises a significant increment of this parameter for MBG-58SGa. On the contrary, pore size remains almost constant for MBG-85S after Ga incorporation.

The chemical compositions of the MBGs are shown in Table 3. The incorporation of $Ga^{3+}$ was carried out based on an ionic exchange with $Ca^{2+}$ and the theoretical values displayed in Table 3 correspond to this assumption. The results obtained by XRF elemental analysis demonstrate that Ga incorporates to the MBGs in a very efficient way. In the case of MBG-58SGa, Ga incorporation is associated with a calcium loosening respect to MBG-58S, in agreement with the assumed mechanism that involved a $Ca^{2+}$ versus $Ga^{3+}$ ionic exchange. On the other hand, Ga incorporation within MBG-85SGa did not result in a significant calcium decrease.

To study the local atomic environment in the MBGs and the effect of Ga incorporation, solid-State $^{29}$Si MAS NMR experiments were performed. Ga free MBGs (Figure 2) show the resonances corresponding to the silicon species $Q^4$, $Q^3$, $Q^2$ and $Q^2_{Ca}$ (with non-bonding oxygens associated to calcium cations). $^1H\rightarrow{^{29}Si}$ CP MAS spectra provide valuable information about the Si environment at the MBGs surface. The $^1H\rightarrow{^{29}Si}$ CP spectrum for MBG-58S decreases the intensity of the $Q^4$ resonances and emphasizes the resonances for $Q^3$ and $Q^2$ due to the presence of Si-OH groups at the MBG surface. The $Q^2$ resonance associated to the presence of $Ca^{2+}$ is emphasized in a higher degree, pointing out that the MBG surface is enriched in $Ca^{2+}$ with respect to the inner part of the MBG walls. On the other hand, MBG-85S does not show $Q^2(Ca^{2+})$ resonance in the single-pulse MAS spectrum and only a weak resonance can be observed in the $^1H\rightarrow{^{29}Si}$ CP experiment. After Ga incorporation, both single pulse and $^1H\rightarrow{^{29}Si}$ CP MAS spectra showed very similar results (Figure S2 in supporting information). The network connectivity of the different MBGs was calculated from the de-convoluted peak areas of single pulse spectra and are shown in Table 4. The results show that MBGs with higher Ca contents (MBG-58S and MBG-58SGa) have lower network connectivity compared to MBG-85S and MBG-85SGa. Moreover, $^{29}$Si MAS NMR experiments indicate that the connectivity of the silica network is not significantly affected after Ga incorporation.

To probe the Ga local environment in the Ga-containing MBGs, $^{71}$Ga solid-state MAS NMR experiments were performed at very high magnetic field (20.0 T). As shown in Figure 3, the experimental $^{71}$Ga central transition (CT) MAS NMR spectra of the MBGs exhibit two resolved resonances with asymmetric line shapes characteristic of a distribution of the quadrupolar interaction. To account for this distribution, the $^{71}$Ga MAS NMR line shapes were reconstructed according to the Gaussian Isotropic Model (GIM) in which the distribution of the electric field gradient is assumed to correspond to a statistical disorder [49,50]. For the two observed resonances, the mean $^{71}$Ga isotropic chemical shifts determined from fits of the experimental spectra are characteristic of respectively six-fold ($\delta_{ISO}$ = 47 ppm) and four-fold ($\delta_{ISO}$ = 172 ppm) coordinated Ga in silicate environments, thus ruling out interaction between Ga ions and calcium phosphate groups[51–53]. This is also supported by the $^{31}$P MAS NMR spectra of the MBGs which remain similar before and after the $Ga^{3+}/Ca^{2+}$ exchange process (see supporting information). For the two MBG compositions investigated, the $^{71}$Ga MAS NMR spectra therefore reveal that the $Ga^{3+}$ cations are mainly incorporated in the silicate network as $GaO_4$ units (about 67 and 73% for MBG-58S and MBG-85S respectively) and that only a weak amount of them forms $GaO_6$ units (about 33 and 27 % for MBG-58S and MBG-85S respectively). This suggests that Ga incorporation into MBG is not only related to a Ga/Ca exchange process but also involves the formation of $GaO_4$ units via Ga-O-Si covalent bonds. It should also be noted that the observed incorporation of $GaO_4$ units in the silicate network requires charge balancing which can be ensured either by $Ca^{2+}$ or $H^+$ cations.

## 3.2 In vitro bioactivity test

Figure 4 shows the $Ca^{2+}$ concentration and pH values of SBF as a function of soaking time of the different MBGs. Calcium levels (Figure 4.a) increase for all the samples during the first 24 h and, the calcium concentration slowly decreases until the end of the assay. Figure 4.b plots the pH evolution of SBF as a function of soaking time, showing a decrease of $H^+$ concentration in SBF because of the ionic exchange with $Ca^{2+}$ from the MBGs.

    The evolution of the MBGs surfaces as a function of soaking time was followed by FTIR spectroscopy through. (Figure S4 in supporting information). Before soaking, the MBGs show a

weak singlet absorption band at 600 cm$^{-1}$ corresponding to the bending vibrational mode of phosphate groups in an amorphous environment, as correspond to $P_2O_5$ included in MBGs compositions. After 6h in SBF, this band increases the intensity and split into a doublet at 560 and 610 cm$^{-1}$. The appearance of this doublet is commonly assigned to the formation of an apatite-like phase when bioactive glasses are in contact with SBF or similar biomimetic solutions [54]. In the case of MBG-58S and MBG-58SGa, an intense band at 1400-1550 cm$^{-1}$ corresponding to the stretching vibrational mode of $CO_3^{2-}$ groups increases as a function of soaking time, indicating that the newly growth phase is a crystalline carbonated hydroxyapatite. This fact is related with the different morphology observed for the apatite crystallites in these samples. SEM observations at higher magnification (supporting information Figure S6) demonstrate that MBG-58S and MBG-58SGa develop an apatite phase made up of agglomerated spherical particles composed by very small crystallites, whereas MBG-85S and MBG85S-Ga form apatite with flake-shape morphology and larger crystallites.

Besides, the presence of a newly formed apatite like phase was confirmed by SEM (Figure 5), indicating that, regardless the Ga presence or absence, the new apatite like phase nucleate and grows on these materials.

### 3.3 Ga in vitro release

Prior to put in contact the cells with the MBGs, we analysed the Ga solubility from the materials in α-MEM (figure S7 in supporting information). After 7 days, Ga(III) concentration in α-MEM reached 803 μM and 71 μM for MBG-58SGa and MBG-85SGa, respectively. After 10 days, the concentrations of Ga released were 860 μM and 86 μM for MBG-58SGa and MBG-85SGa, respectively. These results indicate that, after this period, MBG-58SGa releases the 40% of the its gallium content whereas MBG-85SGa releases only the 12 %.

### 3.4 Effect of Ga on MC3T3-E1 pre-osteoblast like cells

To evaluate Ga effect in osteoblastic bone forming cells, cell culture tests were carried out with MC3T3-E1 pre-osteoblast in the presence and absence of the materials for 1, 4 and 7 days. Fluorescence microscopy images (Figure 6) show that MC3T3-E1 pre-osteoblast cultured in the presence of MBG are flattened and attached to the wells similarly to the cells in absence of materials (supporting information Fig S8). The cells show well-developed actin cytoskeletons organized into long parallel bundles, together with the nuclei in blue stained with DAPI. The purple bodies observed in the images correspond to MBGs particles that absorbed both stains in their mesoporous structure. MBG particles, which are in direct contact with MC3T3-E1 pre-osteoblast (Figure S9 in supporting information), do not lead to any morphological anomaly of the cytoskeleton.

Figure 7.a shows the proliferation of MC3T3-E1 in contact with MBGs. In agreement with the images observed by confocal microscopy, pre-osteoblast cells exhibit a good proliferative behaviour in contact with both, gallium free and gallium containing materials.

The effect of Ga on the MC3T3-E1 activity and early differentiation was studied by measuring the ALP and total protein after 7 days of incubation in contact with the materials (**Figure 7.b**). Significant differences in ALP activity were observed by comparing MBGs with and without Ga. In this case, MC3T3-E1 expressed higher ALP phosphatase activity when they were cultured in the presence of MBG-58SGa and MBG-85SGa, when compared to their Ga-free analogues.

### 3.5 Effect of Ga-doped MBGs on osteoclast cells

First, we measured the effect of the 4 different MBGs on the viability of RAW 264.7 monocytic cells stimulated with RANK-L (Figure 8A). Compared to the control, both MBG-58S and MBG-85S significantly decreased the viability of RAW 264.7 (respectively 27 and 22 %) but, Ga had no supplementary impact.

To evaluate the antiresorptive capability of Ga doped MBGs, the measurement of TRAP expression of mature osteoclasts has been carried out. The evaluation of positive multinucleated stained cells by microscopy let know that Ga dramatically affected the viability of mature OC whereas cells were normal when treated with control MBGs (i.e MBG-58S or MBG-85S). This inhibition is more pronounced in the case of MBG-85SGa. To quantify Ga effect on the osteoclast differentiation, we measured TRAP staining. We have previously demonstrated that TRAP quantification can be considered as a reliable marker of osteoclastogenesis [47]. Indeed, a correlation between the TRAP content per well and the number of TRAP+ multinucleated cells per well has been observed. Figure 8B evidences that the presence of Ga in MBG-58SGa and MBG-85SGa, strongly inhibited the expression of TRAP as compared to control condition.

## 4. Discussion

In this work we propose the incorporation of Ga(III) ions within MBGs as a strategy to improve the bone forming capability of these bioceramics. On the assumption that Ga(III) can inhibit osteoclastogenesis while fostering osteoblasts early differentiation, we envision that Ga-containing MBGs would be excellent bone grafting materials for osteoporotic patients. To understand the potential activity of Ga(III) ions over pre-osteoblasts and osteoclasts, we have carried out a deep physicochemical study, paying special attention to the Ga local environment in Ga-containing MBGs. Our results demonstrate that Ga(III) ions are mainly incorporated as tetrahedral GaO4 units. In this sense Ga (III) would covalently contribute as non-soluble species to the network connectivity of the $SiO_2$-CaO-$P_2O_5$ MBGs. The covalent bonding would be a consequence of the polycondensation of [$GaO_4$] units with the [$SiO_4$] tetrahedrons at the surface of MBGs, favoured by the alkaline conditions during the Ga incorporation process. This mechanism takes place in both MBG-58SGa and MBG-85SGa. Additionally, Ga(III) ions are also incorporated as [$GaO_6$] octahedral species, thus acting as network modifiers and would constitute the main fraction of soluble Ga in MBGs. The incorporation of $GaO_6$ species occurs by means of Ca (II)-Ga(III) ionic exchange with Ca(II) cations of the MBGs. This second incorporation mechanism takes place to a higher extent in MBG-58SGa, as this composition contains more soluble Ca (II) ions acting as network modifiers in MBG-58S. During the Ca(II)-Ga(III) exchange, $Ca^{2+}$ cations are released and the partial dissolution of MBG-58S takes place. Consequently, an enlargement of the pore size occurs, which is reflected in the higher values of pore size distribution observed for MBG-58SGa. In the case of MBG-85SGa, Ca(II)-Ga(III) exchange occurs to a lesser extent due to the smaller amount of Ca(II) available as network modifier. It must be highlighted that the presence of $P_2O_5$ as orthophosphate entraps most of the $Ca^{2+}$ cations in calcium phosphate clusters, as could be observe by $^{31}P$ NMR studies and previously reported by our group [26-29]. In addition, the higher silica network connectivity of MBG-85S avoids the dissolution of this MBG during Ga(III) incorporation and, consequently, the pore size remains almost constant (Table 2). Depending on the composition of the MBG, surface area and porosity change differently after gallium incorporation. In the case of MBG-58SGa, it proceeds by means of calcium-gallium exchange. This mechanism involves the partial dissolution of MBG-58S, thus decreasing the surface area and increasing the pore size. On the other hand, MBG-85SGa evidences a higher decrease of surface area while keeping the pore size after gallium incorporation. These data point out that gallium mainly incorporates on the outer surface of MBG-85S without affecting the inner porous structure by dissolution. These results would agree with those obtained by Ga-NMR that indicate that gallium is mainly incorporate on MBG-85SGa through the polycondensation of $[GaO_4]$ units on the surface.

Despite of the Ca(II) decrease associated to Ga(III) incorporation, Ga-containing MBGs exhibit an excellent *in vitro* bioactive behavior. Both MBG-58SGa and MBG-85SGa developed an apatite like-layer on their surface after 6 hours in contact with SBF. This test, although not definitive, is indicative of the osseointegration capability of this kind of materials under *in vivo* conditions [46, 55]. In this sense, the high surface area and porosity of MBGs would be responsible of the high bioactivity, regardless the calcium loss.

Ga (III) ions do not affect pre-osteoblast proliferation and stimulate the early differentiation toward osteoblast phenotype. The increment of ALP activity of pre-osteoblasts in contact with Ga-containing MBGs respect to Ga-free MBGs indicate that this ion provides signals to stimulate differentiation toward bone forming cells. It must be highlighted that a very small amount of Ga(III) in MBG-85SGa is enough to produce a significant increase of this early differentiation marker respect to MBG-85S. On the other hand, Ga incorporation in MBGs influences TRAP expression suggesting that the presence of Ga disturbs the osteoclastogenesis. These *in vitro* results are in concordance with our previous data relative to the biological effects of Ga in solution and loaded in a calcium phosphate matrix [36,56,57]. Interestingly, the activity of Ga has been closely related with its capability for being released as soluble species. At physiological pH, most of the dissolved Ga is present as gallate, $Ga(OH)_4^-$ although the pH variability in the different bone compartments can modify both, solubility and speciation of Ga. In the case of our MBGs, although Ga solubility is very low (specially for MBG85S-Ga) is enough to inhibit osteoclastogenesis. After ten days in α-MEM, the released gallium reached 70 and 810 μmolar concentrations for MBG85S-Ga and MBG58S-Ga, respectively. Blair et al demonstrated that gallium concentrations between 7-12 μmolar in vivo and around 100 μmolar in vitro, are enough to promote inhibition of osteoclastogenesis [58]. In this sense, both MBGs would be able to supply soluble gallium species to impact on the cultured cells. However, it must be highlighted that osteoclast inhibition is more pronounced in the case of MBG-85SGa, although the amount of Ga released is much lower compared to MBG-58SGa. This fact suggests that gallium bonded at the MBG85S-Ga surface is active, similarly to the experiment of Hall and Chambers, who demonstrated that bonded gallium to bone remain active [32]. In fact, in the case of MBG85S-Ga, gallium is mainly incorporated at the external surface of the MBG85S as indicated by the dramatic reduction of textural properties after Ga incorporation (Table 3). Our results suggest that, in addition to the solubilized gallium, this element would be also incorporated into macrophages in contact with MBG particles, thus inhibiting the differentiation toward osteoclast phenotype.

**Conclusions**

Mesoporous bioactive glasses in the system $SiO_2$-$CaO$-$P_2O_5$-$Ga_2O_3$ have been prepared by impregnation of Ga(III) on two different $SiO_2$-$CaO$-$P_2O_5$. Ga (III) incorporation proceeds according to two different mechanisms: direct condensation within the silica network and ionic exchange with Ca (II), resulting in the formation of network formers $GaO_4$ units and network modifiers $GaO_6$.

The incorporation of Ga within $SiO_2$-$CaO$-$P_2O_5$ MBGs does not inhibit their excellent bioactive behavior.

The presence of Ga in MBGs shows a selective behavior with respect to different cell types. Whereas Ga-doped MBGs stimulate the early differentiation of pre-osteoblast towards osteoblastic phenotype, our experiments also demonstrate that Ga disturbs the osteoclastogenesis. Interestingly, these effects of Ga on bone cells does not require the dissolution of this element.

Considering the excellent bioactive behavior of $SiO_2$-$CaO$-$P_2O_5$-$Ga_2O_3$ based MBGs and their selective behavior with respect to osteoblasts and osteoclasts, these bioceramics might be considered as excellent bone substitutes for regenerative purposes in osteoporotic patients.

**Acknowledgements**

This study was supported by research grants from the Ministerio de Economía y Competitividad (projects MAT2015-64831-R and MAT2016-75611-R AEI/FEDER, UE. M.V.-R. acknowledges funding from the European Research Council (Advanced Grant VERDI; ERC-2015-AdG Proposal 694160).

**Figure captions**

**Figure 1**. Low angle XRD patterns (left) and TEM images of the different MBGs synthesized (right).

**Figure 2**. Solid-state $^{29}$Si single-pulse (left) and cross-polarization (right) MAS NMR spectra of Ga free MBGs.

**Figure 3**: $^{71}$Ga solid state NMR MAS spectra of MBG-58SGa and MBG-85SGa and their deconvolution.

**Figure 4**. Ionic dissolution in different media. (a) $Ca^{2+}$ concentration and (b) pH changes as a function of soaking time in SBF.

**Figure 5**. SEM micrographs of the MBGs before (left) and after soaking in SBF for 24 hours (right)

**Figure 6.** Fluorescence microscopy images of MC3T3-E1 after 4 days in contact with MBGs. Some MBG particles are observed in purple and delimited into a white line. Control image shows MC3T3-E1 after 4 days in the absence of material.

**Figure 7**. (A) Proliferation of MC3T3-E1 pre-osteoblast like cells at 1, 4 and 7 days of culture in contact with the different MBGs synthesized. Statistical significance: *$p < 0.05$ (4 days vs 1 day) #$p < 0.01$ (7 days vs 4 days). (B) ALP activity of MC3T3-E1 pre-osteoblast like cells after 7 days of culture in contact with the different MBGs synthesized. Statistical significance: **$p < 0.05$

**Figure 8.** (A) Viability of Raw 264,7 mouse monocyte cells after 4 days of culture with 20 nM RANKL and in contact with 10 mg/ml of the different MBGs. (B) Effect of MBGs on the TRAP expression of mature OC. Cells were cultured for 4 days in presence of 20 nM RANKL. Then TRAP dosage of cellular extracts was performed, and optical density was read at 562 nm. Results were normalized according to control condition. * $p<0.05$, statistically significant compared to control condition.

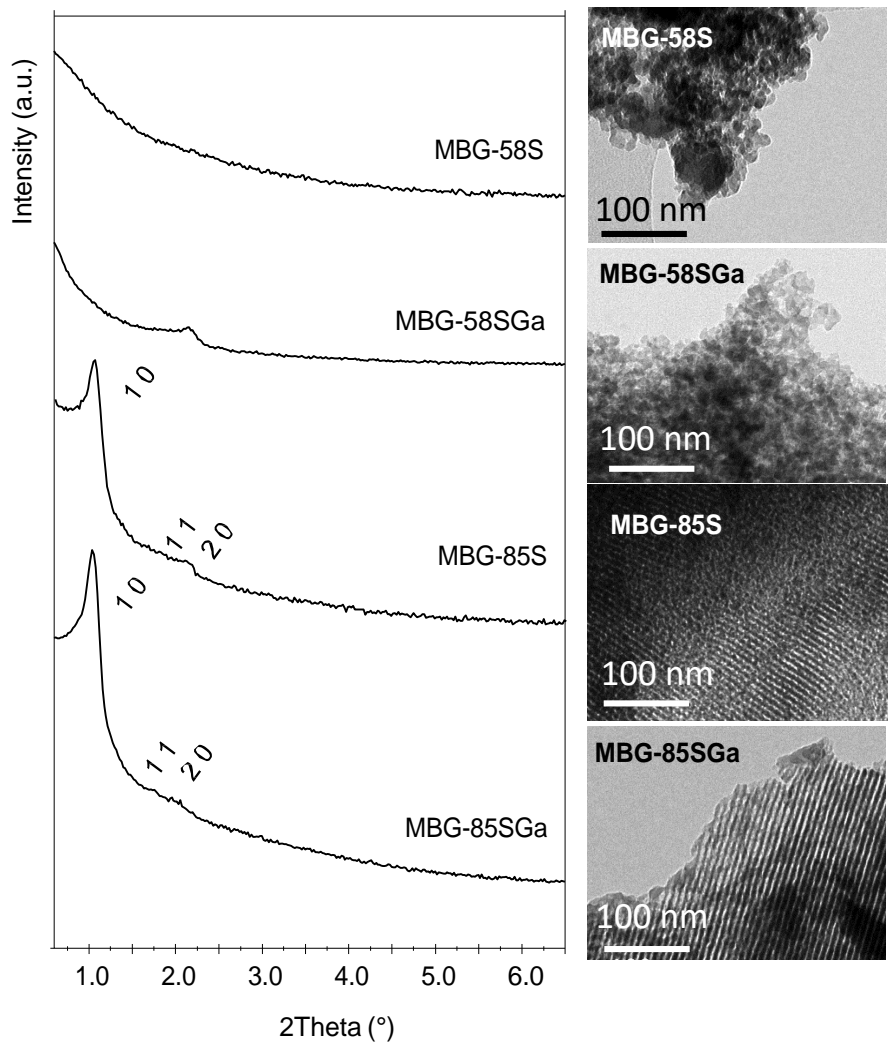

Figure 1

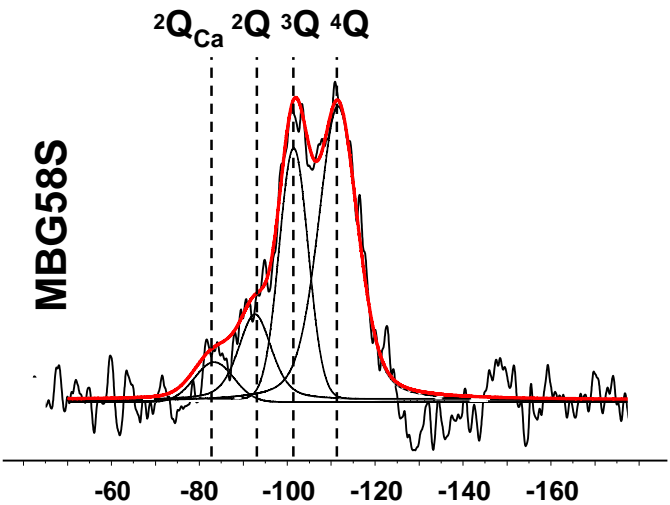
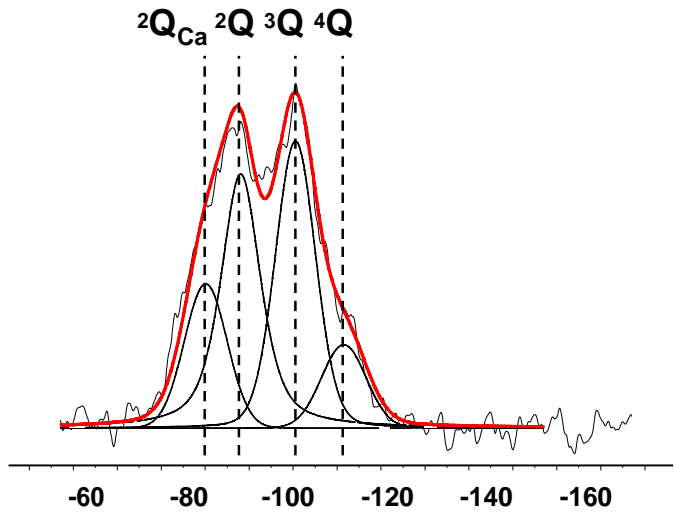
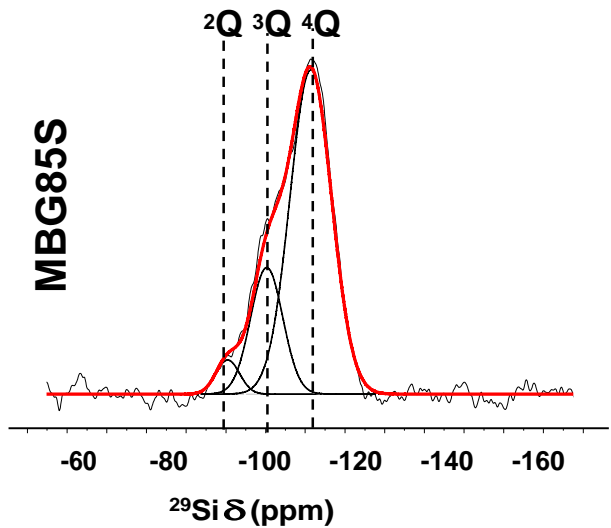
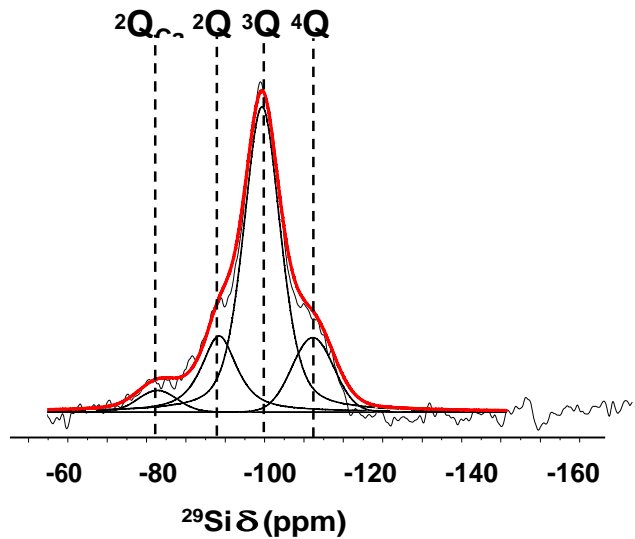

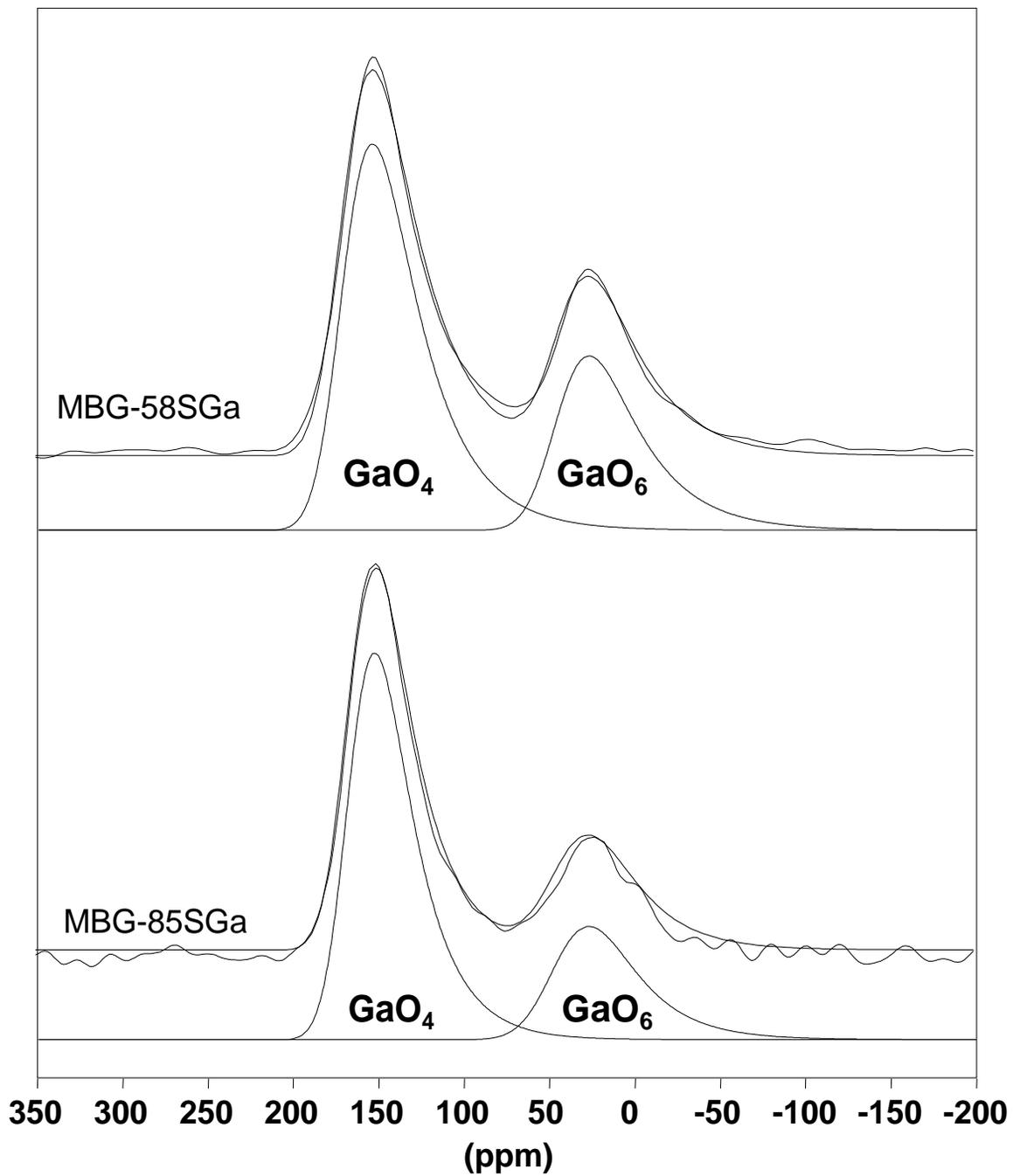

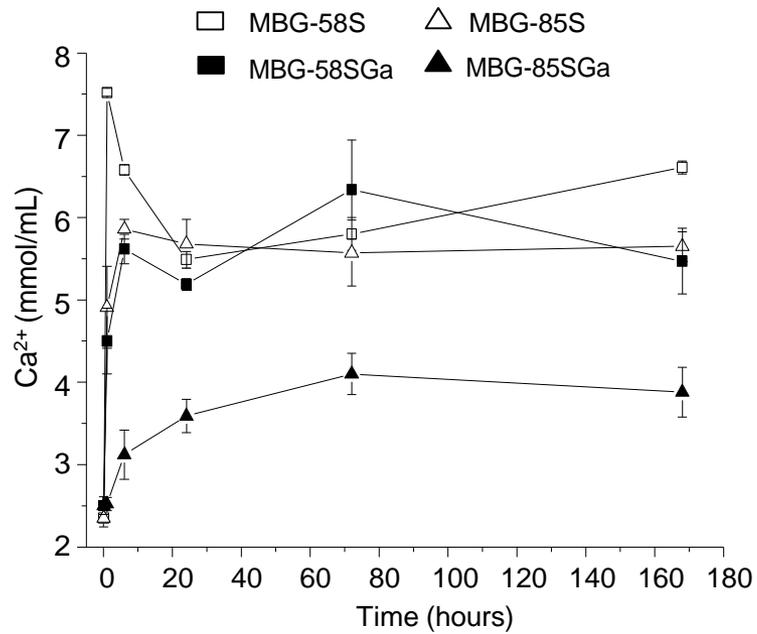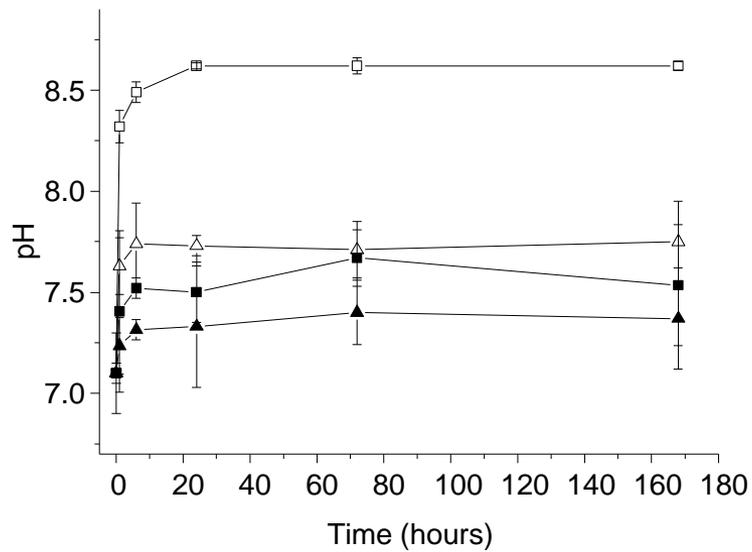

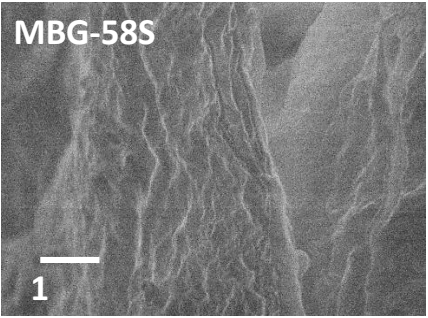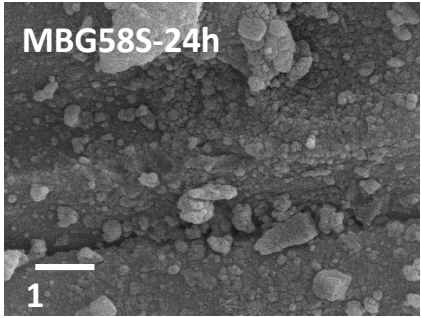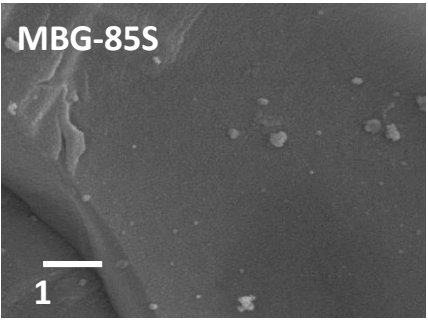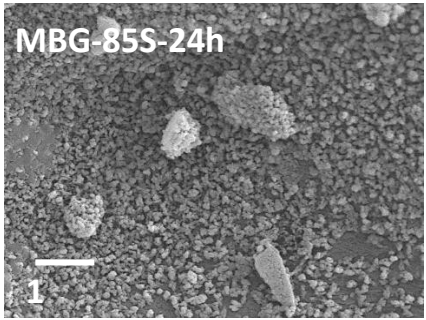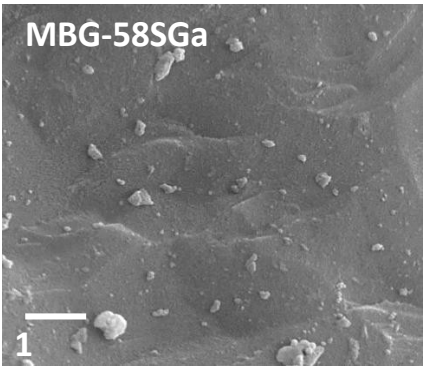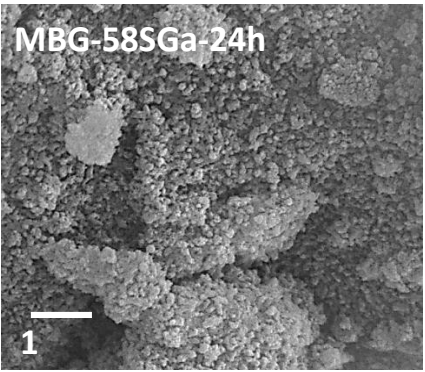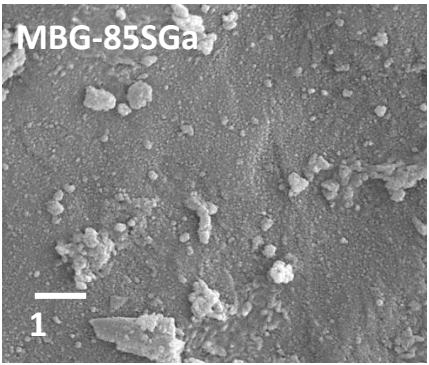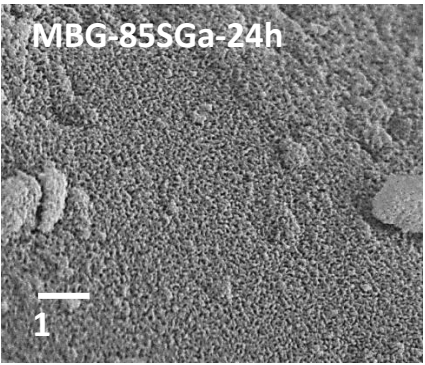

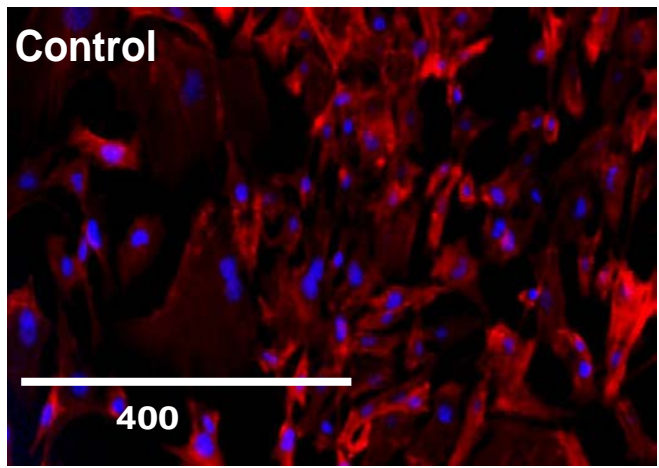
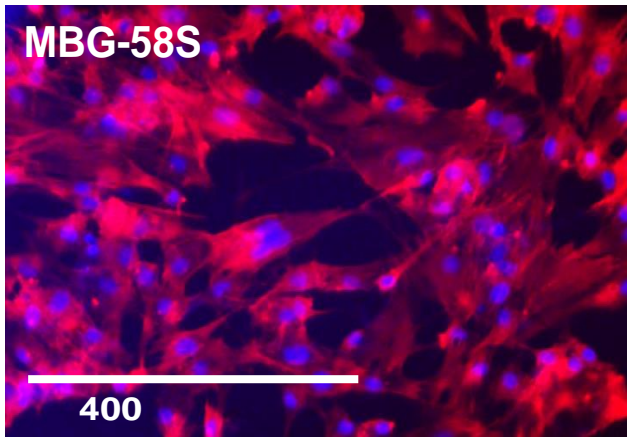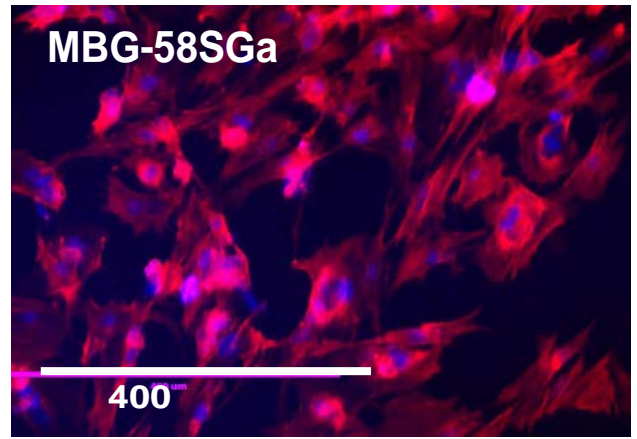
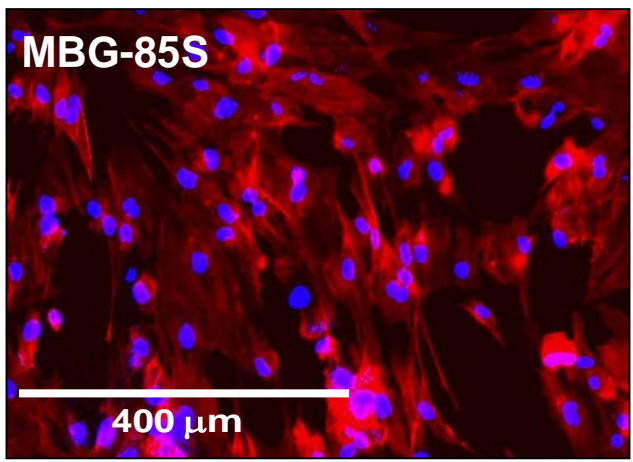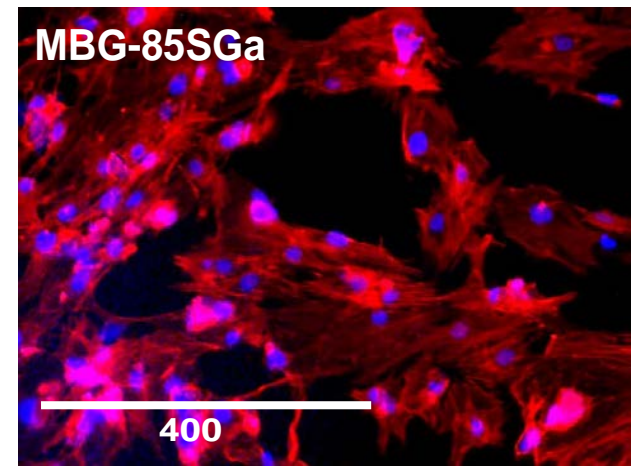

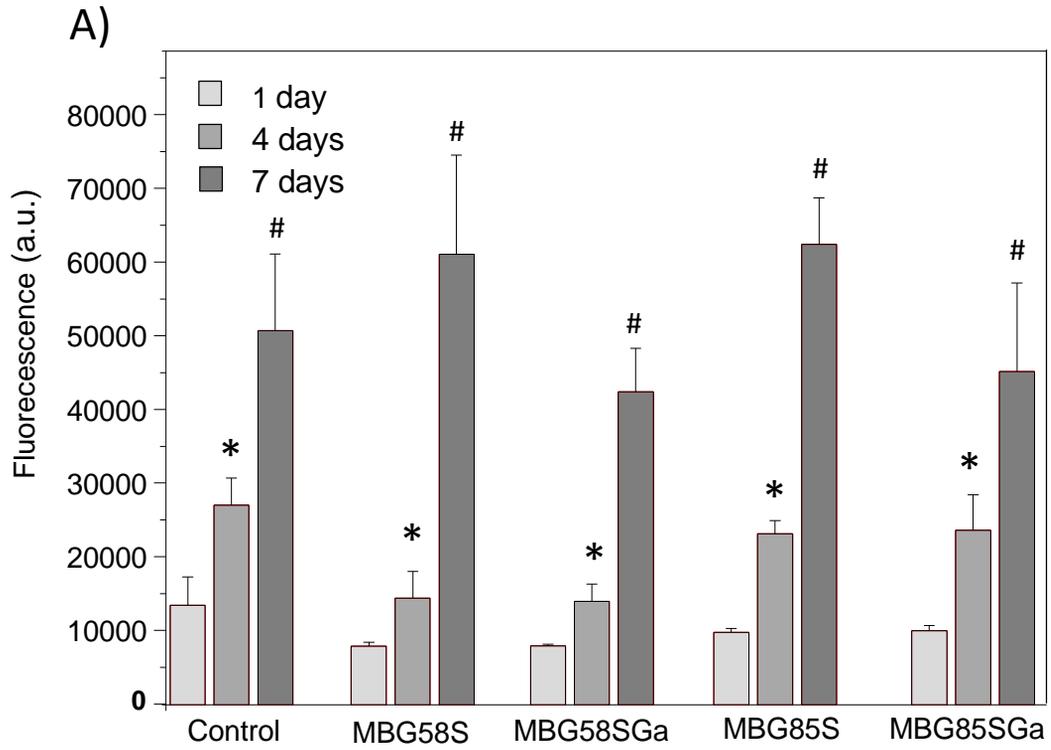
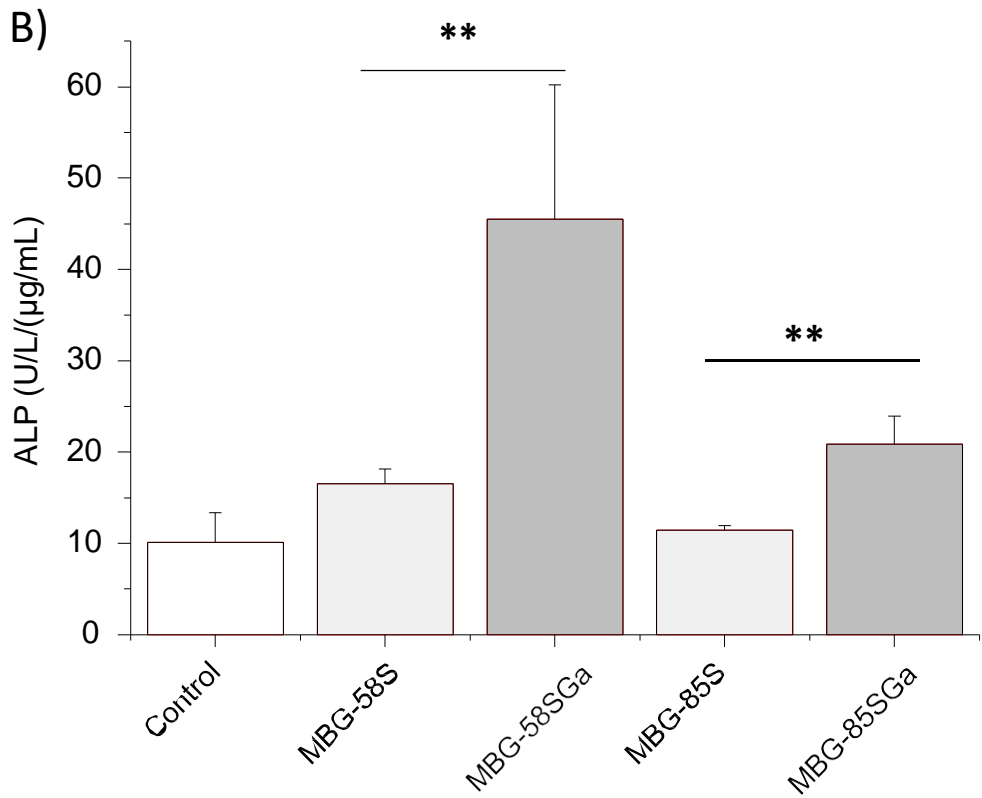